\documentstyle[12pt,graphicx]{article}

\begin{document}

{\it University of Shizuoka}

\hspace*{10.5cm} {\bf US-01-02, v3}\\[-.3in]

\hspace*{10.5cm} {\bf June 2001}\\[-.3in]

\hspace*{10.5cm} {\bf hep-ph/0104226}\\[.3in]

\vspace*{.4in}

\begin{center}

{\Large\bf  Can the Zee Model Explain  \\[.1in]
the Observed Neutrino Data?} \\[.3in]

{\bf Yoshio Koide}\footnote{
E-mail: koide@u-shizuoka-ken.ac.jp} \\

Department of Physics, University of Shizuoka \\ 
52-1 Yada, Shizuoka 422-8526, Japan \\[.1in]

\vspace{.3in}

{\large\bf Abstract}\\[.1in]

\end{center}

\begin{quotation}
The eigenvalues and mixing angles in the Zee model are 
investigated parameter-independently.  When we require
$|\Delta m^2_{12}/\Delta m^2_{23}| \ll 1$ in order to 
understand the solar and atmospheric data simultaneously,
the only solution is one which gives bi-maximal mixing.
It is pointed out that the present best-fit value of 
$\sin^2 2\theta_{solar}$ in the MSW LMA solution
cannot be explained within the framework of the Zee model,
because we derive a severe constraint on the value of
$\sin^2 2 \theta_{solar}$, 
$\sin^2 2 \theta_{solar} \geq 1 -(1/16)
(\Delta m^2_{solar}/\Delta m^2_{atm})^2$.
\end{quotation}

\vfill
\newpage

Of the neutrino mass matrix models proposed currently,
the Zee model\cite{Zee} is a very attractive one,
because the model can naturally leads to a large neutrino 
mixing with few parameters \cite{Petcov,Smirnov-Tanimoto,Jarlskog}.
The neutrino mass matrix $M_\nu$ in the basis on which
the charged lepton mass matrix $M_e$ is diagonal is given 
by the form
$$
M_\nu \simeq m_0 
\left(\begin{array}{ccc}
0 & a & c \\
a & 0 & b \\
c & b & 0 
\end{array}\right) \ ,
\eqno(1)
$$
where
$$
a= f_{e\mu} (m_\mu^2 -m_e^2) \ ,
$$
$$
b= f_{\mu \tau} (m_\tau^2 -m_\mu^2) \ ,
\eqno(2)
$$
$$
c= f_{\tau e} (m_e^2 -m_\tau^2) \ ,
$$
and $f_{e\mu}$, $f_{\mu \tau}$ and $f_{\tau e}$ are 
lepton-number violating Yukawa coupling constants with
the Zee scalar $h^+$.
It is known that if we consider a Zee mass matrix with
$a=c\gg |b|$, the model can give a nearly bi-maximal
mixing \cite{Petcov,Jarlskog,Ghosal}
$$
U_\nu =
\left(\begin{array}{ccc}
\cos\theta & -\sin\theta & 0 \\
{\frac{1}{\sqrt{2}}}\sin\theta & {\frac{1}{\sqrt{2}}}\cos\theta & 
-{\frac{1}{\sqrt{2}}} \\
{\frac{1}{\sqrt{2}}}\sin\theta & {\frac{1}{\sqrt{2}}}\cos\theta 
& {\frac{1}{\sqrt{2}}} 
\end{array}\right),
\eqno(3)
$$
where
$$
\tan\theta=
\sqrt{-m_{\nu 1}/m_{\nu 2}} \ ,
\eqno(4)
$$
$$
\Delta m_{12}^2 = m^2_{\nu 1}-m^2_{\nu 2} 
\simeq 2 \sqrt{2}  ab \ , \ \ 
\Delta m_{23}^2 = m^2_{\nu 2}-m^2_{\nu 3} \simeq 2 a^2 \ ,
\eqno(5)
$$
which lead to 
$$
{\frac
{{\Delta}m_{12}^2}
{{\Delta}m_{23}^2}}
\simeq{\sqrt{2}}
{\frac{b}{a}} \ .
\eqno(6)
$$
Furthermore, if we assume a badly broken horizontal symmetry
SU(3)$_H$ and we put a simple ansatz on the transition matrix
elements in the infinite momentum frame 
(not on the mass matrix), we can obtain the relations
\cite{Koide-Ghosal}
$$
f_{ij}= \varepsilon_{ijk}\frac{m_k^e}{m_i^e +m_j^e} f\ , 
\eqno(7)
$$
where $f$ is a common factor and $m_i^e=(m_e,m_\mu,m_\tau)$, 
so that we can predict
$$
{\frac
{{\Delta}m_{12}^2}
{{\Delta}m_{23}^2}}
\simeq \sqrt{2}
\frac{m_e}{m_\mu} = 6.7\times10^{-3} \ ,
\eqno(8)
$$
which is in excellent agreement with the observed value 
(best fit value) \cite{atm,Garcia}
$$
\left(
{\frac{{\Delta}m_{solar}^2}
{{\Delta}m_{atm}^2}}
\right)_{exp}\simeq
{\frac
{2.2\times10^{-5}\ {\rm eV}^2}
{3.2\times10^{-3}\ {\rm eV}^2}}
= 6.9\times10^{-3} \ .
\eqno(9)
$$

Thus, the Zee model is very attractive from the
phenomenological point of view.
However, most authors who investigated the Zee
neutrino mass matrix have failed to give the
observed value $\sin^2 2\theta_{solar} \simeq 0.7$ 
in the MSW LMA solution \cite{Bahcall}, 
although it is
easy to give the bi-maximal mixing (3).
It is a serious problem for the Zee model
whether the model can fit the observed value
$\sin^2 2\theta_{solar} \simeq 0.7$ or not.
In the present paper,
from a parameter-independent study of the
Zee neutrino mass matrix (1),
we conclude that the value of 
$\sin^2 2\theta_{solar}$ must satisfy a severe 
constraint 
$\sin^2 2 \theta_{solar} \geq 1 -(1/16)
(\Delta m^2_{solar}/\Delta m^2_{atm})^2$
in the Zee model with 
$\Delta m^2_{solar}/\Delta m^2_{atm} \ll 1$.
The similar subject has also been discussed by
Frampton and Glashow \cite{Frampton}.
However, the constraint obtained in the present
paper is more explicit and very severe.
This constraint will force  us to abandon the Zee model
or to modify the original Zee model to an extended
version with some additional terms.

The mass matrix (1) is diagonalized by a unitary matrix
$U_\nu$ as
$$
U_\nu^T M_\nu U_\nu = D_\nu \equiv {\rm diag}
(m_1, m_2, m_3) \ .
\eqno(10)
$$
The Maki-Nakagawa-Sakata (MNS) \cite{MNS} matrix $U_{MNS}$
is given by $U_{MNS}=U_\nu$, because the charged lepton
mass matrix is diagonal in the Zee model.
In order to obtain the relations among the mass matrix
parameters and the mass eigenvalues, we define the
Hermitian matrix $H_\nu$ as
$$
H_\nu = M_\nu^\dagger M_\nu \ ,
\eqno(11)
$$
so that we obtain
$$
U_\nu^\dagger H_\nu U_\nu = D_\nu^* D_\nu = {\rm diag}
(|m_1|^2, |m_2|^2, |m_3|^2) \ .
\eqno(12)
$$
The form of $H_\nu$ is explicitly given by
$$
H_\nu = H_0 - H_1 \ ,
\eqno(13)
$$
where
$$
H_0 = m_0^2 (|a|^2 +|b|^2 +|c|^2) {\bf 1} \ ,
\eqno(14)
$$
$$
H_1= m_0^2 \left(
\begin{array}{ccc}
|b|^2 & -c^* b & -a^* b \\
-b^* c & |c|^2 & -a^* c \\
-b^* a & -c^* a & |a|^2 
\end{array} \right) \ ,
\eqno(15)
$$
and ${\bf 1}$ is a $3\times 3$ unit matrix.
The matrix $H_1$ is diagonalized as
$$
U_\nu^\dagger H_1 U_\nu = {\rm diag}
(h_1, h_2, h_3) \ ,
\eqno(16)
$$
and the eigenvalues $h_i$ satisfy the equation
$$
h_i^3 -(|a|^2 +|b|^2 +|c|^2) m_0^2 h_i^2 +
4 |a|^2 |b|^2 |c|^2 m_0^6 = 0 \ .
\eqno(17)
$$
By re-defining $m_0$, without losing generality,
we can take $|a|^2 +|b|^2 +|c|^2 =1$, so that
the solutions $h_i= m_0^2 x_i$ are described
only by one parameter 
$$
|q|^2=  |a|^2 |b|^2 |c|^2 \ ,
\eqno(18)
$$
as
$$
 x_i^3 - x_i^2 + 4 |q|^2 = 0 \ .
\eqno(19)
$$
The equation (19) has three real solutions $x_i$ 
only when $|q|^2 < 1/27$.
The behaviors of the solutions $x_i$ are illustrated
in Fig.~1.
The mass squared $|m_i|^2$ is given by
$$
|m_i|^2 =(1 -x_i) m_0^2 \ .
\eqno(20)
$$
{}From Fig.~1, we find that the cases which can explain
the observed fact $|\Delta m^2_{12}/\Delta m^2_{23}| \ll 1$
are only the cases with $|q|^2 \simeq 1/27$ and 
$|q|^2 \simeq 0$.

\vspace{.2in}
\begin{figure}
\begin{center}

\includegraphics[width=8cm]{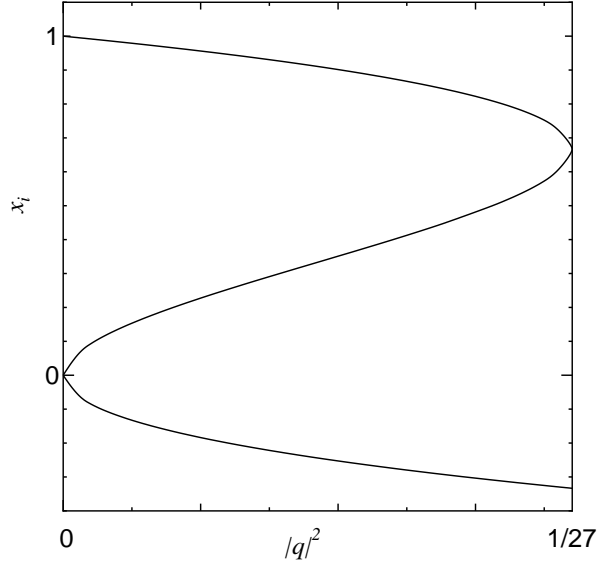}

\end{center}
\caption{
The eigenvalues $x_i$ $(i=1,2,3)$ versus $|q|^2$.
The solutions $x_i$ of the equation (19) have
real three values only in the range 
$0 \leq |q|^2 \leq 1/27$.
The values $x_i$ take $(0,0,1)$ and $(-1/3,2/3,2/3)$
at $|q|^2=0$ and $|q|^2=1/27$, respectively.
The mass eigenvalues $|m_i|^2$ are given by
$|m_i|^2 =(1-x_i)m_0^2$.
}
\label{k_xi}
\end{figure}
\vspace{.2in}


For the case with $|q|^2 \simeq 1/27$, by putting
$$
x_1 = \frac{2}{3} +\varepsilon_1  \ , \ \ 
x_2 = \frac{2}{3} -\varepsilon_2 \ , \ \ 
x_3 = -\frac{1}{3} +\varepsilon_3\ , \ \ 
|q|^2 = \frac{1}{27} - \varepsilon_q^2 \ ,
\eqno(21)
$$
and by putting (21) into the equation (19), we can obtain
$$
\varepsilon_1 \simeq \varepsilon_2 \simeq 2\varepsilon_q 
\ , \ \ 
\varepsilon_3 \simeq 4 \varepsilon_q^2 \ . 
\eqno(22)
$$
so that we obtain
$$
\frac{\Delta m^2_{21}}{\Delta m^2_{32}} \simeq 4
\varepsilon_q \ .
\eqno(23)
$$
On the other hand, from Eq.~(16), we obtain
$$
(H_1/m_0^2)_{ii} = |U_{\nu i1}|^2 x_1 +
|U_{\nu i2}|^2 x_2 + |U_{\nu i3}|^2 x_3 \ .
\eqno(24)
$$
For the case with $|q|^2 \simeq 1/27$, Eq.~(24)
gives
$$
(H_1/m_0^2)_{ii} \simeq \frac{2}{3} -|U_{\nu i3}|^2
+ 2 \varepsilon_q (|U_{\nu i1}|^2 -|U_{\nu i2}|^2)
\ ,
\eqno(25)
$$
i.e.,
$$
|b|^2 \simeq \frac{2}{3} -|U_{\nu 13}|^2 \ , \ \ 
|c|^2 \simeq \frac{2}{3} -|U_{\nu 23}|^2 \ , \ \ 
|a|^2 \simeq \frac{2}{3} -|U_{\nu 33}|^2 \ . 
\eqno(26)
$$ 
Since we know that the only solution under the
conditions $|a|^2 +|b|^2 +|c|^2=1$ and 
$ |a|^2 |b|^2 |c|^2 \simeq 1/27$ is
$|a|^2 \simeq  |b|^2 \simeq |c|^2 \simeq 1/3$,
the relations (25) yield
$$
|U_{\nu 13}|^2 \simeq \frac{1}{3} \ , \ \ 
|U_{\nu 23}|^2 \simeq \frac{1}{3} \ , \ \ 
|U_{\nu 33}|^2 \simeq \frac{1}{3} \ , 
\eqno(27)
$$
which give
$$
\sin^2 2\theta_{atm}=4 |U_{\nu 23}|^2 |U_{\nu 33}|^2
\simeq \frac{4}{9} \ .
\eqno(28)
$$
The value (28) is too small to explain the observed 
value \cite{atm} $\sin^2 2\theta_{atm} \simeq 1.0$, 
so that the case with $|q|^2 \simeq 1/27$ is ruled out.

Next, we investigate the case with $|q|^2\simeq 0$.
By putting
$$
x_1 = -\varepsilon_1  \ , \ \ 
x_2 = \varepsilon_2 \ , \ \ 
x_3 = 1 -\varepsilon_3 \ , 
\eqno(29)
$$
and by putting (29) into the equation (19), we can obtain
$$
\varepsilon_1 \simeq 2 |q|(1-|q|) \ , \ \  
\varepsilon_2 \simeq 2 |q|(1+|q|) \ , \ \  
\varepsilon_3 \simeq 4 |q|^2 \ . 
\eqno(30)
$$
so that we obtain
$$
\Delta m^2_{12} \simeq 4|q| m_0^2 \ , \ \ 
\Delta m^2_{23} \simeq (1-2|q|^2) m_0^2 \ ,
\eqno(31)
$$
$$
\frac{\Delta m^2_{12}}{\Delta m^2_{23}} \simeq 4
|q| \ .
\eqno(32)
$$
On the other hand, form the relation (24), we
obtain
$$
(H_1/m_0^2)_{ii} \simeq |U_{\nu i3}|^2
- 2 |q| (|U_{\nu i1}|^2 -|U_{\nu i2}|^2)
+2|q|^2 (1-3|U_{\nu i3}|^2)
\ ,
\eqno(33)
$$
so that we obtain
$$
(H_1/m_0^2)_{22}=|c|^2 \simeq |U_{\nu 23}|^2 \ , \ \ 
(H_1/m_0^2)_{33}=|a|^2 \simeq |U_{\nu 33}|^2 \ , 
\eqno(34)
$$
and
$$
\sin^2 2\theta_{atm} \simeq 4|a|^2 |c|^2 \ .
\eqno(35)
$$
Generally, the only solution of the equation 
$xy\simeq 1/4$ for the positive numbers $x$ and
$y$ under the condition $x+y<1$ is
$x \simeq y \simeq 1/2$.
Therefore, the solution of the equation
$\sin^2 2\theta_{atm}= 4 |U_{\nu 23}|^2 |U_{\nu 33}|^2 \simeq 1$
under the condition $|U_{\nu 23}|^2 +|U_{\nu 33}|^2 
= 1-|U_{\nu 13}|^2 <1$ is
$$
|U_{\nu 23}|^2 \simeq |U_{\nu 33}|^2 \simeq \frac{1}{2}
\ , \ \ |U_{\nu 13}|^2 \simeq 0 \ ,
\eqno(36)
$$
and also the solution of the equation
$\sin^2 2\theta_{atm} = 4 |a|^2 |c|^2 \simeq 1$ 
under the condition $|a|^2 +|c|^2=1 -|b|^2 <1$ is
$$
|a|^2 \simeq  |c|^2 \simeq \frac{1}{2} \ , \ \ 
|b|^2 \simeq 0 \ .
\eqno(37)
$$
The result (37) means
$$
|q|^2 \simeq \frac{1}{4}|b|^2 \ .
\eqno(38)
$$

The $(1,1)$ component of the equation (33) gives
$$
 |b|^2 \simeq |U_{\nu 13}|^2
-  |b| (|U_{\nu 11}|^2 -|U_{\nu 12}|^2)
+ \frac{1}{2}|b|^2 (1-3|U_{\nu 13}|^2)
\ .
\eqno(39)
$$
When we put
$$
|U_{\nu 11}| = \sqrt{1-|U_{\nu 13}|^2} \cos \theta \ ,
\ \ 
|U_{\nu 12}| = \sqrt{1-|U_{\nu 13}|^2} \sin \theta \ ,
\eqno(40)
$$
we obtain
$$
\sin^2 2\theta_{solar} \simeq \sin^2 2\theta
\simeq 1 - \frac{1}{4} |b|^2 \left( 1 -
2 \frac{|U_{\nu 13}|^2}{|b|^2}
\right)^2 \ ,
\eqno(41)
$$
where
$$
|b| \simeq \frac{1}{2} 
\frac{\Delta m_{solar}^2}{\Delta m_{atm}^2} \ .
\eqno(42)
$$
A model which gives $|U_{\nu 13}|^2 =0$ cannot
obviously give a sizable deviation from
$\sin^2 2\theta_{solar}=1$.
However, if $|U_{\nu 13}|^2 \sim |b|$, then
the value of $\sin^2 2\theta_{solar}$ is 
sensitive to the value of $|U_{\nu 13}|^2$.
Therefore, we must estimate the  value of 
$|U_{\nu 13}|^2$ carefully.

We use the relations
$$
\sum_{k=1}^3 (H_1/m_0^2)_{ik} U_{\nu kj} = U_{\nu ij} x_j \ .
\eqno(43)
$$
For $j=3$, we obtain
$$
|b|^2 U_{\nu 13} - c^* b U_{\nu 23} - a^* b U_{\nu 33}
= U_{\nu 13} x_3 \ ,
\eqno(44)
$$
$$
-b^* c U_{\nu 13} +|c|^2 b U_{\nu 23} - a^* c U_{\nu 33}
= U_{\nu 23} x_3 \ ,
\eqno(45)
$$
$$
-b^* a U_{\nu 13} - c^* a U_{\nu 23} +|a|^2 U_{\nu 33}
= U_{\nu 33} x_3 \ .
\eqno(46)
$$
By eliminating $U_{\nu 23}$, we obtain the relation
without any approximation
$$
U_{\nu 13} = \frac{-2 (x_3-1+|b|^2)b a^* U_{\nu 33}}{
(|a|^2-|c|^2) |b|^2 +(x_3-|b|^2)(x_3-1+|b|^2)} \ .
\eqno(47)
$$
If we use the approximate expression $x_3\simeq
1-4|q|^2 \simeq 1-|b|^2$, the factor 
$(x_3-1+|b|^2)$ becomes vanishing.
Therefore, in order to estimate the factor
$(x_3 -1+|b|^2)$ more precisely, 
we use the following expression of $x_3$ to the 
order of $|q|^4$,
$$
x_3 \simeq 1 -4|q|^2 (1+8|q|^2) \ .
\eqno(48)
$$
Then, we can show
$$
x_3 -1 +|b|^2 \simeq |b|^2 \left[
(|a|^2 -|c|^2)^2 + 7|b|^4 \right] \ .
\eqno(49)
$$
Since we know that $|b|^2$ is very small value,
i.e., $|b|^2 \simeq (1/4)(\Delta m^2_{solar}/
\Delta m^2_{atm})^2$, we investigate only the
case $(|a|^2 -|c|^2)^2 \geq |q|^4$.
Then, from Eq.~(47), we can obtain
$$
U_{\nu 13} \simeq -2(|a|^2 -|c|^2) b a^* U_{\nu 33}
\ ,
\eqno(50)
$$
i.e.,
$$
|U_{\nu 13}|^2 \simeq (|a|^2 -|c|^2)^2 |b|^2
\ .
\eqno(51)
$$
On the other hand, we can show that the quantities
$(\Delta m^2_{12}/\Delta m^2_{23})^2$ and
$\sin^2 2\theta_{atm}=4|U_{\nu 23}|^2 |U_{\nu 33}|^2$
are insensitive to the parameter $(|a|^2 -|c|^2)$.
Therefore, from Eqs.(47) and (51), we can obtain
the following parameter-independent relation
$$
\sin^2 2\theta_{solar} \simeq 
1 -\frac{1}{4} \left[1 -2(|a|^2 -|c|^2)^2 \right]^2 
|b|^2 \geq 1 - \frac{1}{16} \left( 
\frac{\Delta m^2_{solar}}{\Delta m^2_{atm}}\right)^2
\ ,
\eqno(52)
$$
where we have used $[1 -2(|a|^2 -|c|^2)^2]^2 \leq 1$.

The constraint (52) cannot be loosened even if we
consider the renormalization group equation (RGE) effects.
The mass matrix form (1) is given by the radiative diagrams
at the low energy scale, where the charged lepton mass
matrix is given by the diagonal form.
Although the coupling constants $f_{ij}$ given in Eq.~(2) are
affected by the RGE, since our conclusion (52) is independent of
the explicit values of the parameters $a$, $b$ and $c$
in Eq.~(1), the conclusion (52) cannot be loosen even
by taking RGE effects into consideration.

However, we must note that the mass matrix form (1)
based on only the one-loop radiative mass diagrams.
When we take two-loop diagrams into consideration,
as pointed out by Chang and Zee \cite{Chang-Zee},
non-vanishing contributions appear in the diagonal 
elements of $M_\nu$.
For the case which gives $\sin^2 2\theta_{atom} \simeq 1$,
the relations (37) are required, so that the relations
$|f_{e \mu}| m_\mu^2 \simeq |f_{e \tau}| m_\tau^2
\gg |f_{\mu \tau}| m_\tau^2$ are required.
Then, as discussed in Ref.~\cite{Chang-Zee},
we can estimate
$$
|M_{ \nu 12}| \simeq |M_{ \nu 13}| \gg |M_{ \nu 23}| 
> |M_{ \nu 11}| \simeq  |M_{ \nu 22}| \gg |M_{ \nu 33}| 
\ , 
\eqno(53)
$$
where $|M_{ \nu ij}| \propto f_{ij}(m_i^2 -m_j^2)$ and
$|M_{ \nu ii}| \propto |f_{12}||f_{23}||f_{31}|(m_j^2 -m_k^2)$ 
($j \neq i \neq k$).
We interest in a value of the ratio $|M_{\nu 11}/M_{\nu 23}|$.
If the ratio is negligibly small, the result (52) will be
still valid, but if the ratio is sizable, then the result
(52) will be valid no more.
According to Ref.~\cite{Chang-Zee}, we estimate
$|M_{\nu 11}/M_{\nu 23}|$ as
$$
\left| \frac{M_{\nu 11}}{M_{\nu 23}} \right| \sim
\frac{|f_{e \mu}||f_{\mu \tau}||f_{\tau e}|}{
16\pi^2 |f_{\mu \tau}|} \simeq 
\frac{|f_{e \mu}|^2}{16\pi^2} \left( \frac{m_\mu}{m_\tau}
\right)^2 < 10^{-5} \ .
\eqno(54)
$$
Therefore, we conclude that the severe constraint (52) is 
still valid even if we take two-loop diagrams into
consideration.

However, note that if the mass matrix (1) is not 
due to the Zee mechanism, but due to a seesaw mechanism, 
$M_\nu \simeq - m_L M_R^{-1} m_L^T$, the form of $M_\nu$
will be changed by the RGE effects. 

Therefore, we can conclude that when we require
$\Delta m^2_{solar}/\Delta m^2_{atm} \ll 1$  for
the Zee model, although we can give 
$\sin^2 2 \theta_{atm}\simeq 1$, but, at the
same time,  
the value of $\sin^2 2 \theta_{solar}$ must also
be highly close to one.
On the other hand, in contrast to the theoretical bound
(52), the best fit value of $\sin^2 2 \theta_{solar}$
is
$$
\sin^2 2 \theta_{solar} \simeq 0.66 \ ,
\eqno(55)
$$
for the MSW LMA solution \cite{Bahcall}.
The prediction $\sin^2 2 \theta_{solar}\simeq 1.0$
is in poor agreement with the observed data
(in the outside of the region 99\% C.L.).
Of course, the value (55) is a best-fit value,
and it does not mean that the Zee model is
ruled out.
However, if the data in future exclude the value
$\sin^2 2 \theta_{solar}\simeq 1.0$ completely, 
we will be forced
to abandon the Zee model, at least, for the MSW
LMA solution.
At present, if we still adhere to the Zee model,
the only solution which we should take is 
the Just So$^2$ solution \cite{Raghavan} with 
$\sin^2 2 \theta_{solar}\simeq 1.0$.
However, the Just So$^2$ solution does  not always
the best one of the possible candidates (the 
best fit solutions) at present
(for example, the MSW LMA solution gives 
$\chi_{min}^2=29.0$, while the Just So$^2$ solution
$\chi_{min}^2=36.1$ \cite{Bahcall}).

In conclusion, we have investigated the Zee neutrino
mass matrix (1) parameter-independently.
When we require that the value
$\Delta m^2_{solar}/\Delta m^2_{atm} =
\Delta m^2_{12}/\Delta m^2_{23}$ should be very
small, the possible solutions are only two
cases with $|q|^2\equiv |a|^2 |b|^2 |c|^2 \simeq
1/27$ and $|q|^2 \simeq 0$ where 
$|a|^2+|b|^2+|c|^2=1$.
The case with $|q|^2\simeq 1/27$ leads to 
$\sin^2 2 \theta_{atm}\simeq 4/9$, so that
the case is ruled out.
The case with $|q|^2\simeq 0$ leads not only to 
$\sin^2 2 \theta_{atm}\simeq 1$, but also to
$\sin^2 2 \theta_{solar} \geq 1 -(1/16)
(\Delta m^2_{solar}/\Delta m^2_{atm})^2$.
The prediction $ \sin^2 2 \theta_{solar} \simeq 1.0$
is in poor agreement with the observed data.
However, in spite of such a problem, 
the Zee model is still attractive to us, 
because the model can naturally 
lead to a nearly bi-maximal mixing.
Therefore, we would like to expect that the problem
will be overcome by some future modification of 
the original Zee model.
For examples, the following attempts will be promising:
introducing a new doubly charged scalar $k^{++}$ 
in order to obtain sizable two-loop contributions
\cite{k++}, and
introducing right-handed neutrinos in order to 
additional mass terms,
and embedding the original Zee model into an $R$-parity
violating SUSY model \cite{R-viol} and into an
$R$-parity conserving SUSY model \cite{R-consv}, 
and so on.

\vspace*{.4in}

\centerline{\Large\bf Acknowledgments}

The author would like to thank  A.~Ghosal and
H.~Fusaoka for their helpful discussions and comments,
and Midori Kobayasi for informing a mathematical
theorem in uniqueness of the solutions. 
He also thank S.~T.~Petcov for informative discussion 
and valuable comments.
Furthermore, he is also grateful to M.~Yasue for valuable
discussion on the two-loop effects in the Zee model, 
to A.~Zee for his quick response to an inquiry about
the two-loop effects,  and
to M.~Tanimoto and J.~Sato for helpful discussions on SUSY 
versions of the Zee model. 


\end{document}